# Reconstruction of ancient conceptual landscapes in the Nile Valley

**Giulio Magli**

Faculty of Civil Architecture, Politecnico di Milano.
Piazza Leonardo da Vinci 32, 20133 Milan, Italy. E-Mail: Giulio.Magli@Polimi.it

**Abstract:** Conceptual landscapes in Egypt show a remarkable continuity – for as long as 2000 years – in the use of symbols and in the interplay between natural and man-built features. *Directionality*, both in the sense of succession of elements and of orientation of single buildings and tombs, plays a key role in governing the landscape in accordance with the idea of "cosmic" order, which was the basis of the temporal power of the pharaoh. Comparing satellite image with local surveys and using simple web-based instruments for tracing visibility lines helps in understanding connections and messages which were meant to be clear and obvious in ancient times but may be lost, or forgotten, today. In particular, the prominent role of astronomical and topographical alignments in the planning of successive monuments comes out at sites like Abydos, Giza, Dahshur and at Western Thebes. The way in which the same symbols and elements were elaborated by the "heretic" pharaoh Akhenaten in planning the landscape of his capital at Amarna is also highlighted.

**Keywords:** Ancient Egypt, Ancient Landscapes, Web-based Archaeo-topography

## 1. Introduction

Archaeoastronomy, the science exploring the relationship of ancient architecture with the sky, is a relatively new discipline. It started from the pioneering efforts of Gerald Hawkins and Alexander Thom in the sixties of the last century. These authors were the first to put in evidence in a systematic way the possible relevance of astronomical alignments in the interpretation of ancient sites. Their work was biased however by several drawbacks, the main one being the fact that any Archaeoastronomical analysis must be inter-disciplinary, and must take in full account the historical context and the archaeological records. Since then, Archaeoastronomy evolved in as a comprehensive, multi-disciplinary science [1,2]. At the same time, also Archaeology has "evolved". In particular, an increasing importance has been acquired by *cognitive* aspects [3]. The cognitive-science approach to archaeological remains, or cognitive archeology, can be defined - according to Colin Renfrew - as the study of past ways of thought as inferred from material remains. As such, it clearly involves the relationship between ancient art, science and mind and the way in which the ancient thought and lore were embodied in such things as the plan of temples, monuments, but also *entire landscapes.* From this point of view, the archaeoastronomical analysis can be considered in a broader context as the "sky-counterpart" of the analysis of such ancient *conceptual* landscapes.

An ancient civilization where conceptual landscapes can be seen "in action" with an impressive continuity in the course of more than two millennia is Egypt [4]. Cognitive aspects of the landscape in Egypt have been studied by Lehner [5], Jeffreys [6] and Romer [7]. In recent years, the present author studied several issues related to the topography of conceptual landscapes in the case of the Old and the Middle Kingdom pyramids' fields, taking into a special account the aspects related to the ancient sky as a "cognitive" element [8,9,10]. This analysis benefited very much of the use (admittedly at a quite elementary level) of web-based applications such as Google Earth. In fact, although it goes without saying that nothing can substitute the direct experience of an ancient site (and indeed, the present author did survey the sites mentioned in the present paper) some general features of the ancient landscapes can be more clearly recognized from satellite images, especially if such images are integrated with simple but powerful tools like the ruler of Google Earth, which allows an immediate estimate of distances and azimuths. This is due to the fact that elements not existing when the landscape under exam was conceived can alter the perception of it today, or preclude the view between inter-related elements. In other cases, the ruined state of some buildings has the consequence that they are not visible or barely visible, while they were prominent elements of the horizon in ancient times. Finally, and this is a fundamental point in the case of Egypt, ancient *inter-visibility* was in many cases much greater than today, due to modern pollution and/or the presence of later buildings. Of course, the roundness of the earth has always to be taken into account, but this is easily done with the so-called horizon formula, which states that from an height of h meters the visibility is well approximated by the value of $\sqrt{13h}$ in Kilometers (so that for a person 2 meters high the visible horizon is something more than 5 Kilometers). Once this control has been done, and therefore *theoretical* inter-visibility between distant sites is established, to try to understand today the possible role of inter-visibility and the possible existence of mutual alignments one is not only invited, but forced to use the above mentioned tools. Usually, the error of projection introduced by programs like Google Earth remains small on the distances (at most a few tens of Kilometers) which come into play [11]. For instance, the pyramids of Giza were conceived to be visible from the place of the ancient temple of Heliopolis, located 24 Kms far on the opposite bank of the Nile. The exciting experience of the Great Pyramid, the last surviving wonder of the world, occupying the far horizon to the west was still enjoyable from Heliopolis in the 19th century, as shown by old photos and pictures. Today, there is no choice to live such an experience, and not only due to to the buildings in between, but due to the haze and pollution of the modern city of Cairo, which strongly limit visibility. However, Google Earth allows us to test the incredible precision and seriousness of the ancient architects (see section 3 below).

I present here the main ideas inspiring ancient conceptual landscapes in the Nile Valley, stressing in each case their astronomical and topographical connections and showing how these ancient connections can be better understood today using a web-based application. Most of the paper is an overview, but several issues are original and presented here for the first time, especially in Section 5, devoted to the funerary landscape of Western Thebes in the New Kingdom.

## 2. The earliest conceptual landscape in Egypt: Abydos.

Egypt (excluding the Delta area) is a very peculiar place from the environmental point of view: a short strip of fertile terrain – the "gift of the Nile" - inundated once a year and crossed by the river in (roughly) south to north direction. This livable area is surrounded by desert on both sides. Desert commences at the end of the cultivated land. Usually, very near to this border, both to the east and the west, the valley rises up in relatively high stone outcrops, crossed by dried rivers or *Wadi*.

If the Nile flows essentially south-north, giving the first "main axis" to human life, the Sun travels east-west giving the second one. The places to the west, where the sun sets and starts – in the Egyptian conception - his perilous journey into the hours of the night, were associated since very early times with places of death and rebirth. Indeed it is here, on the west bank of the Nile, that some of the most spectacular and complex conceptual landscapes of humanity were conceived and constructed. The landscapes we are going to speak about are therefore, in a sense, funerary landscapes, since they are all connected with the afterlife. Interestingly however, rebirth was also associated to the north. Indeed a fundamental component of the funerary beliefs was in the "rebirth" of the pharaoh in a appropriate place together with the "imperishable" stars, namely the circumpolar stars, which are visible every night of the year. As a consequence, this led to a almost maniacal precision in the cardinal alignment of the pyramids, especially during the Old Kingdom.

It is customary to divide the history of Egypt into wide periods [12]. The early-dynastic period starts with the unification of the state around 3100 BC. It is followed by the Old Kingdom (2686 BC – 2181 BC circa). After a phase of political fragmentation and anarchy (First Intermediate Period) the country is again unified around year 2040 BC by the pharaoh Mentuhotep when, formally, begins the Middle Kingdom (circa 2040-1790 BC). The Middle Kingdom terminates with the 13[th] Dynasty and the invasion of the Country by the Hyksos. The regaining of the sovereignty, started in Thebes, happens with the 18[th] Dynasty and the beginning of the New Kingdom (circa 1550-1150 B.C.)

The first man-modeled conceptual landscape was conceived during the early-dynastic period at Abydos [4]. Abydos was already one of the most important holy places in Egypt; later, starting from the Middle Kingdom, it became the main cult center of Osiris, and in the New Kingdom the pharaoh Seti I constructed here one of the most magnificent temples of the country. We are mostly interested here in the early-dynastic royal Necropolis which was founded in the desert valley today called Umm el Quaab. All the kings of the First Dynasty, and two of the Second, were buried in subterranean tombs at this site. The funerary landscape is composed by the following main elements [13]:

- A building, located near the cultivation limit, devoted to the cult of the deceased king. These buildings are huge open-air rectangular structures of mudbricks, surrounded by graves, most probably of sacrificed persons. They were erased (ritually demolished) after the death of the king, so that only the foundations are being recovered by archaeologists. However, the last one constructed, that of king Khasekhemwy, is still standing. Usually called Shunet el-Zebib, is an imposing monument, with walls still raising up to 11 meters, and sides of 137 and 76 m .

**Figure 1.** The funerary landscape of king Khasekhemwy at Abydos
(1) Royal enclosure (2) Processional way (3) Royal tomb (4) Wadi (5) Mountain of Anubis.
The red line denotes the meridian (image courtesy of Google Earth).

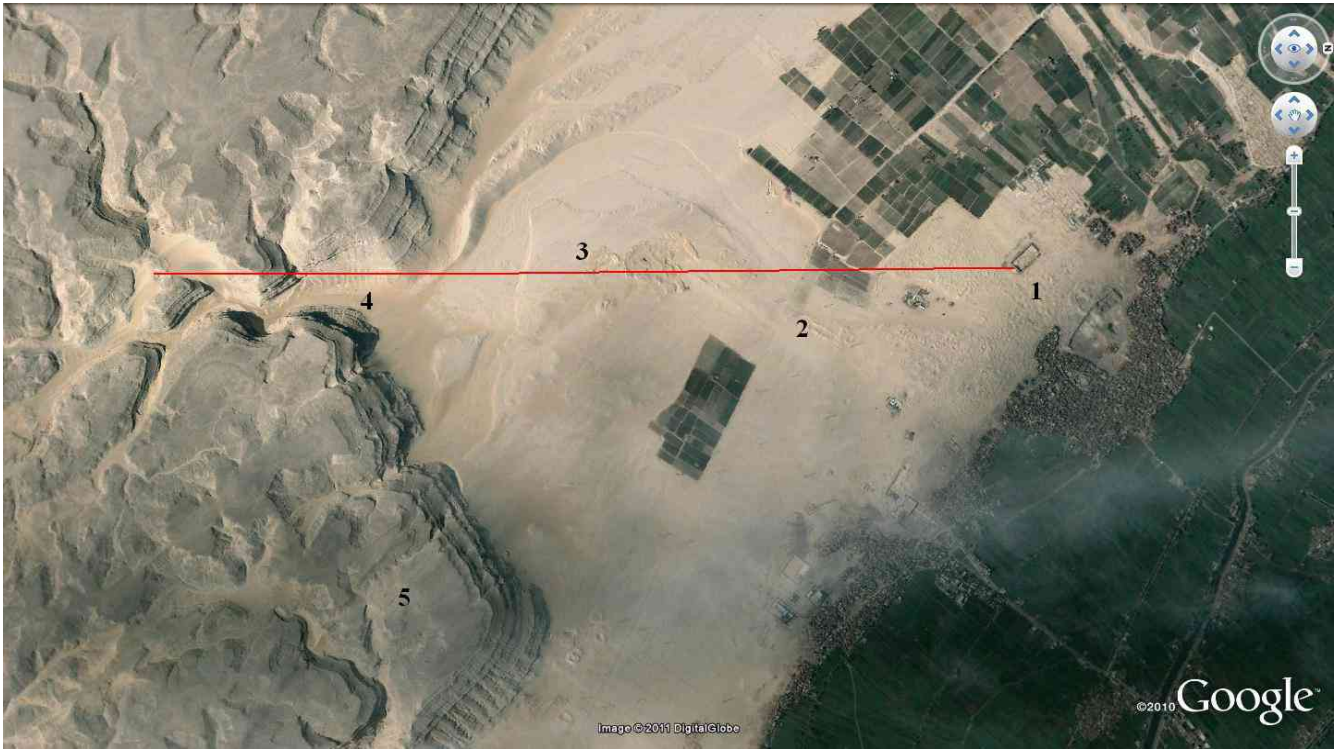

- a processional (unpaved) pathway some 1.5 Kms long, leading from the "cult" funerary area to the Necropolis of the royal tombs
- the subterranean royal tomb, perhaps surmounted by a small mound
- a wadi (and therefore a natural element which influenced the choice of the site) located in the hills due south of the Necropolis. The prominent hill to the east of the wadi was sacred to Anubis, and pharaoh Senwostret III (section 4) later constructed his tomb below it.

A fundamental point in the project of all the subsequent monuments of this area (enclosures and tombs) is orientation. First of all, the enclosures "proceed to the south-west" in the sense that subsequent monuments were usually planned to the south-west of the preceding ones; the same holds – in a even more crystal-clear way - for the tombs. Further, both the tombs and the enclosures have a clear tendency to be oriented along inter-cardinal directions. For instance, the Shunet el-Zebib has azimuth 44° (longest sides). All the other enclosures and tombs have similar orientations. Recently, it has been proposed that this peculiar pattern of orientations aroused from the the will of "mediating" between the meridian alignment to the north and the direction orthogonal to the Nile [14]. I tend rather to see it as a specific, regional pattern specific to Abydos – perhaps originally imported from contemporary Mesopotamian temples, or inspired by the rising and setting positions of the Milky Way in that period - to which all later structures conformed for many centuries (including the Osireion and the Sethi I temple). Other temples and tombs received similar orientations in different places of Egypt as well; however, the presence of "local" patterns of orientation can hardly be negated. For instance, after Khasekhemwy, the kings of the second dynasty moved from Abydos to Saqqara for their tombs. At Saqqara –where all pyramids and pyramid temples will be orientated cardinally later on – their tombs are orientated cardinally.

Since the tombs at Umm el Quaab are interred and are not a visible feature of the landscape, to have a clear look at the way the original Khasekhemwy complex was conceived we can resort to Google Earth. It then becomes immediately clear the geometry which inspired it. The main elements of the complex are seen to form an "ordered" landscape: the processional pathway runs along the meridian, while the enclosure and the tomb at its two ends are oriented inter-cardinally. The wadi is actually due south of the complex; therefore, there is no specific solar reference in its presence, since of course the sun was – and is – seen on the wadi only at culmination, and never at rising or setting. Rather, the Wadi may have functioned as a symbol of a "gate of the afterworld"[4].

## 3. Cognitive aspects of the Memphite sacred landscape

With the sacred landscape in Abydos we encountered a first example of ordered landscape. The "cosmic order", or *Maat*, was a key element of ancient Egyptians thought and beliefs. It was a complex interplay between a solar-stellar religion and the contemporary presence of the living God, the pharaoh himself. The king was the keeper of the cosmic order on the earth and was later doomed to live in eternity - as well as his dynastic ancestors before him - together with the circumpolar stars and the Sun God. These concepts became effective with the Old Kingdom and the Age of the Pyramids, and are visible on the ground both in the project of such extraordinary monuments and in their topographical arrangement. A full discussion of the sacred landscape during the "Age of the pyramids" would be, however, far beyond the limits of the present work. I will limit myself to discuss briefly the most famous and important case, which of course is the Giza Necropolis.

Construction of the pyramids begins with king Djoser, the builder of the Step Pyramid in Saqqara. The short phase of the step pyramids ends with the beginning of the $4^{th}$ dynasty and Khufu's father Snefru (see next section). With the $4^{th}$ dynasty the pyramidal complexes acquire a definitive scheme which will remain unchanged up to the end of the Old Kingdom [15,16]. The scheme is again composed by three main elements, this time all built in stone. Indeed, the enclosure is substituted by a "Valley Temple" and the processional pathway becomes a monumental causeway which leads to the pyramid-tomb. On the east front of the pyramid another temple is built. Again, orientation is fundamental; this time the main direction runs east-west, while orientation of buildings is rigorously cardinal, with the sides of pyramids and temples skillfully parallel to the cardinal directions. In particular, entrance corridors of the pyramids were always directed to the north, and therefore to the circumpolar stars, while the front sides of the temples were directed to the east. In such a context, with the $4^{th}$ dynasty the cult of the Sun becomes more and more relevant; in particular the Sphinx of Giza, near the Valley Temple of Khafre, is oriented due east as well and was associated with the king's funerary complex as a solar symbol [17].

**Figure 2.** The Necropolis of Giza with the "Giza axis" highlighted (image courtesy of Google Earth).

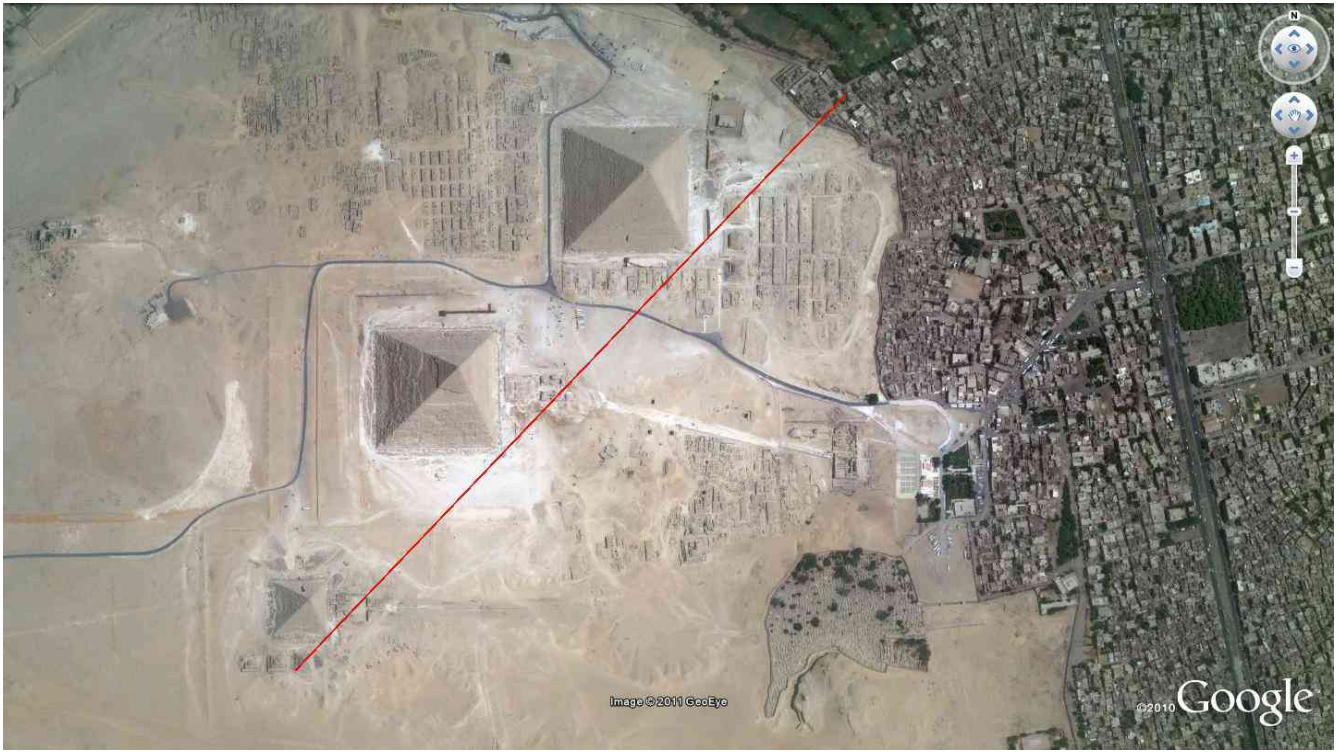

**Figure 3.** The "Giza axis" highlighted and prolonged up to the area of the Temple of Heliopolis. The red line is about 24 Km long (image courtesy of Google Earth).

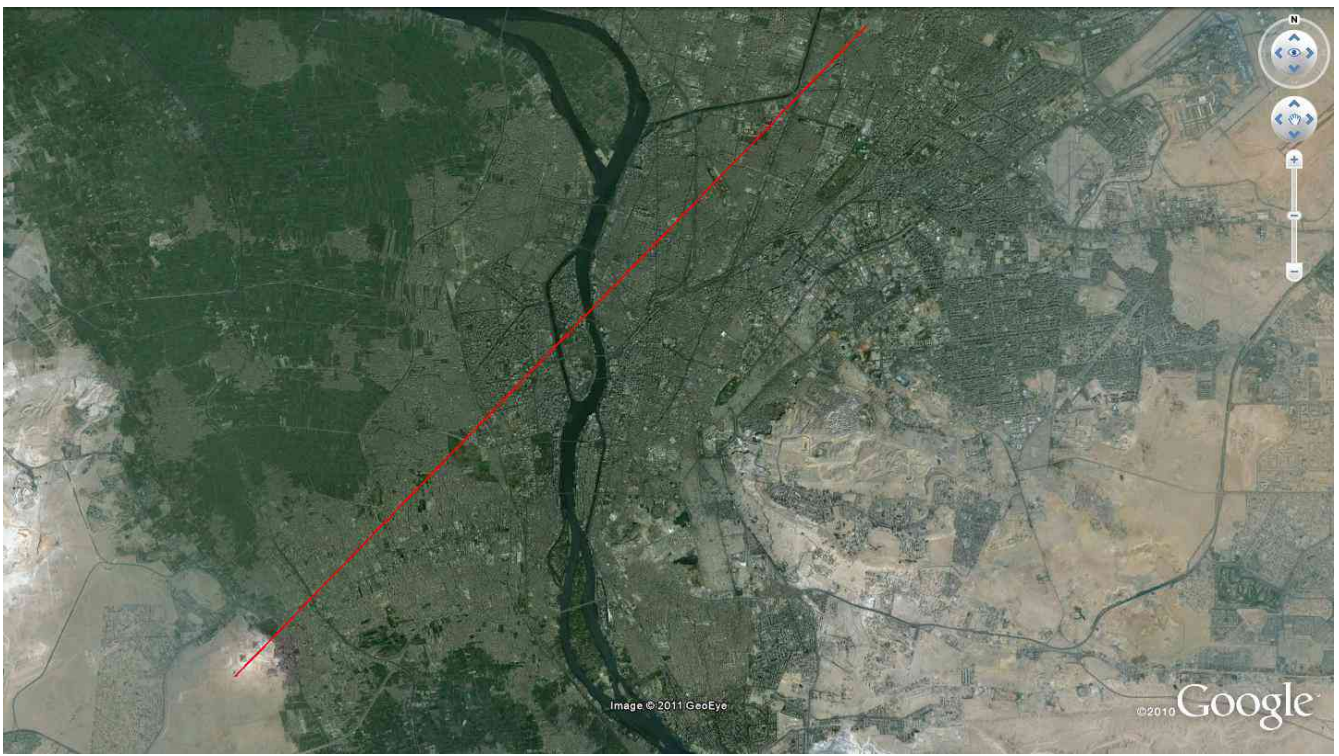

The pyramidal complexes were built in "clusters" or "pyramid fields", at Giza (4th dynasty), Abusir (5th dynasty) and Saqqara (6th dynasty) respectively. These are desert plateaus located not far from the capital Memphis; Giza and Abusir were essentially virgin soil, while Saqqara was already the revered place of Djoser's tomb. The main "cognitive" feature at Giza (and also at Abusir which will not be discussed here) is a imposing topographical axis, which governed the successive planning of the pharaoh's pyramids and their complexes [5]. The axis runs across the south-east corners of the pyramids and ideally connects the monuments with the temple of Heliopolis on the opposite bank of the Nile. Such a temple was the fundamental center of the Sun cult, and therefore the alignment helps to "declare the affinity" of the rulers with the Sun God [4]. The monuments align one after the other in such a way that, if seen from Heliopolis, they "fade" each other in a sort of sacred mirage [18]. It is important to put in evidence that such an effect – and its meaning - were not hidden, or concealed in a sort of alleged esoteric legacy: quite the reverse, the pharaohs wished to make their ideas and the origins of their power as explicit and concrete as possible in the planning of their funerary complexes. However, it is impossible *for us today* to catch visually this message, because of pollution, haze, and interposed buildings. Therefore, a simple and powerful instrument like Google Earth can again be used.

Using Google Earth, the legendary accuracy and seriousness of the builders of the Giza pyramids comes out in a striking way: the Giza diagonal, when prolonged across the Nile, passes some tens of meters from the obelisk of Heliopolis, which is standing very near the entrance of the (today destroyed) ancient temple. The line is *not* a visibility line for a standing person since Heliopolis is quite far from Giza (about 24 km). Yet the horizon formula shows that a sign-post - say 20 meters tall - placed in Heliopolis would have been visible from Giza, and of course the reverse was true - once and forever - as soon as the construction of the Giza pyramids reached a similar height.

**4. The Middle Kingdom pyramids**

After the first intermediate period, Egypt entered in a new phase of unity and renaissance, the so called Middle Kingdom. The pharaohs started again to construct huge pyramidal complexes, whose remains are visible on the ridge of the desert between Dahshur and the northern rim of the Fayoum oasis. Although these remains cannot be compared with the magnificent stone pyramids of the Old Kingdom, also the Middle Kingdom pyramids were conceived and built to be a visible symbol of power and to convey a series of messages related to the divine nature of the kings and their dynastic rights to kingship. Actually, the culture and religion of the Middle Kingdom appears to be a mixture of innovation and "archaism" [19]. From the point of view of the symbolic aspects of the landscape, the most important message can be individuated in a recall to the kingship of Snefru, the father of Khufu This connection was made *explicit* on the ground by means of a rather complex system of topographical, symbolic connections with the Snefru monuments [20].

Snefru (2600 BC circa) constructed two couples of monuments. The most famous ones are the magnificent pyramids – each one more than 100 meters high - located in Dahshur (South Saqqara) called today the Bent and the Red Pyramid (the Bent Pyramid owes its name to a sudden softening of its inclination, which was effected when the construction had reached 49 meters, while the Red

Pyramid owes its modern name to the reddish hue of the limestone used to build it). These paired monuments were perhaps meant to represent a symbolic horizon, the hieroglyph sign 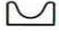 associated with afterlife, when viewed from people ascending to the Saqqara plateau. At Snefru times this path was indeed free up to the entrance of the Djoser complex on its right and a person ascending the plateau would have seen the two giant pyramids of Dahshur standing alone at the profile of the horizon (actually, this impressive experience is still enjoyable today, especially in clear days). Perhaps the two artificial mountains of Snefru actually represented the two (re-united) parts of the country themselves. As a matter of fact, immediately thereafter, the son and first of the "solar" kings, Khufu, will design his funerary project at Giza following the same pattern, but adding to the „paired mountains" the sun setting in between, corresponding to the hieroglyph *Akhet* 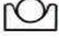. This was achieved with a spectacular hierophany, which can still be experienced from the Sphinx area at the summer solstice, when the sun setting between the two giant pyramids replicates one time a year in the sky the very same name of the Great Pyramid, *Akhet Khufu* or "Horizon of Khufu" [5,18,21].

The other double project of Snefru (most probably conceived before in temporal order, since the Meidum pyramid was perhaps only completed by Snefru) is located about 40 Kms south of Dahshur and composed by the Meidum pyramid and the so-called Seila pyramid. Meidum was perhaps initially conceived as a Step Pyramid, but resembles today a sort of huge tower. Her "companion" is a small step pyramid located on the hills to the west. This "minor" pyramid is located on a somewhat prominent desert outcrop, overlooking the Fayoum oasis. It has no interior structure (so it is not a tomb) and, at least in the opinion of who writes, its role was of ideal companion of Meidum, which is much greater but located in the flat land, as "outpost cenotaphs" signaling the royal power in the approach to the capital, some 60 Kms further north. The two monuments indeed appear to be strictly related, since they are only 10 Kms apart and located approximately on the same parallel. Today they are barely inter-visible with the naked eye – so that again a satellite view is of much help to put them into their correct mutual relationship - however in ancient times visibility was certainly much better.

The four Snefru monuments were standing already from more than 600 years when - after Amenemhet I and Senwostret I who built their pyramids close to the new capital established at Lisht - Amenemhet II was the first to return to Dahshur. The pyramid of this king is almost completely destroyed, but the trace of its enclosure is readily seen on satellite images. It is then seen that the complex was located in a carefully chosen position with respect to the Snefru complex; in particular, the line of the south base of the Red Pyramid, when prolonged due east, intersects a dense area of 4th dynasty tombs which could not be moved. The north side of the "temenos" (perimeter) wall of the Amenemhet II pyramid is located immediately to the south of these tombs. In this way, the complex was positioned to obtain a perspective effect with the much higher, but farthest in the desert, Red Pyramid of Snefru.

The son of Amenemhet II, Senwostret II, constructed his pyramid complex at El-Lahun, in the Fayoum oasis. The second pyramid newly constructed at Dahshur, that of Senwosret III, was planned to the north of that of Amenemhet II. Again, the project was carefully placed taking into account the existing monuments and creating topographical relationships. First of all, a meridian (North-south) line ideally connects the west side of the temenos wall of Senwosret III and the front (east) side of the temenos wall of Amenemhet II.

**Figure 4.** The Necropolis of Dahshur (1) Bent Pyramid (2) Red Pyramid (3) Amenemhet II (4) Senwosret III (5) Amenemhet III. Causeways of the Bent Pyramid and of the Senwostret III pyramids and geometrical connections between monuments highlighted (image courtesy of Google Earth).

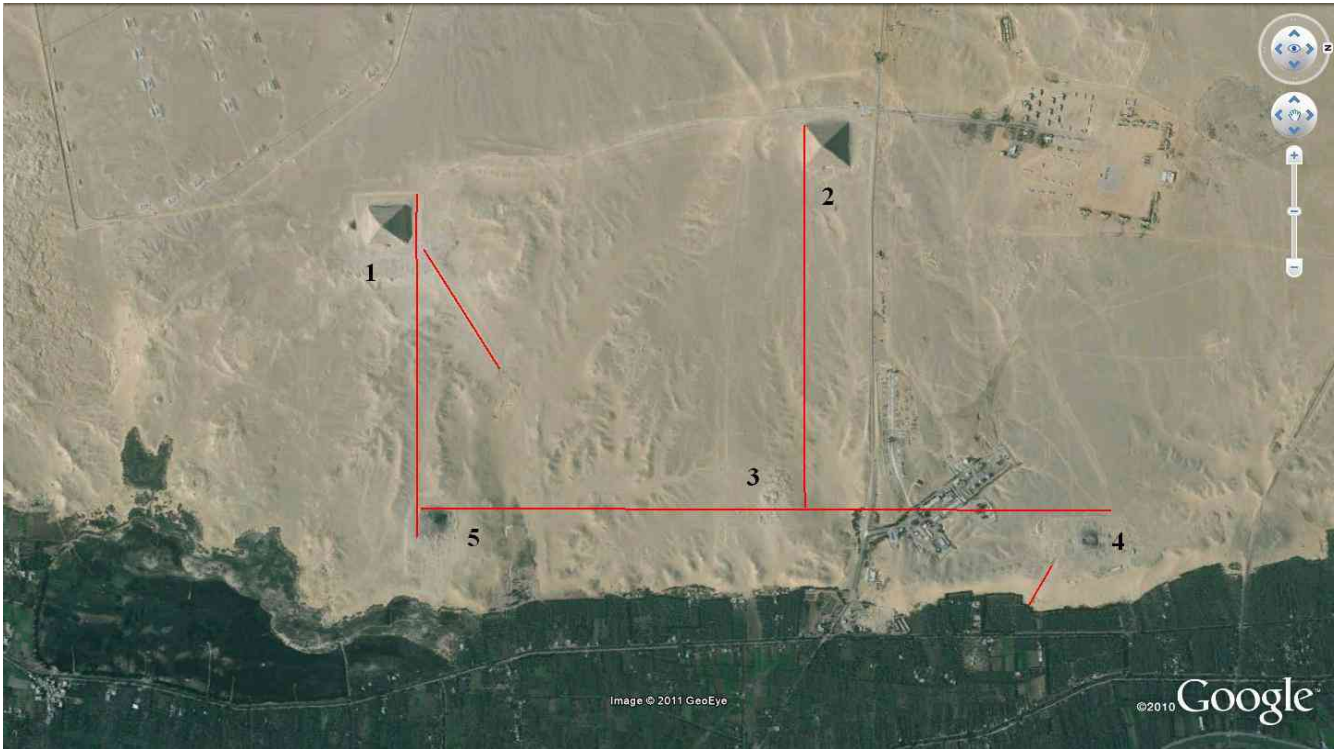

Further, a symmetry relationship between Senwosret III and the already existing projects is introduced by the orientation of the causeway of the complex. Indeed the causeway of the Bent Pyramid - the southernmost pyramid at Dahshur - is oriented (from the Valley Temple to the pyramid) at 240°. This means that, for an observer looking along the causeway, the Sun at the winter solstice was seen to disappear behind the huge mole of the pyramid [22]. The architects who designed the causeway of the Senwosret III complex – the northernmost pyramid of Dahshur - choose to create a configuration symmetrical to that designed more than 600 years before for Snefru, since it points to the setting sun at the summer solstice. Interestingly, there is the strong possibility that the Senwosret III pyramid was actually only a cenotaph, and that this king was buried in a funerary complex he constructed at Abydos [19]. In this case the parallel to the Bent Pyramid, likely a cenotaph as well, would be strengthened.

Also the last Middle Kingdom pyramid built at Dahshur, that of Senwostret III' son Amenemhet III, was perhaps conceived as a cenotaph. The project of this monument - today called Black Pyramid – was again designed carefully to take into account the existing ones in order to harmonize the new element in the human-made landscape and thus to keep Maat, the Cosmic Order, in the already old royal Necropolis. First of all, the existing meridian was taken into account: it runs indeed along the west side base of the Black Pyramid. To fix the position of the pyramid along the meridian, the project took into account the position of the Bent Pyramid to the west, and again the new pyramid was planned in order to create a perspective effect between the new and the old one.

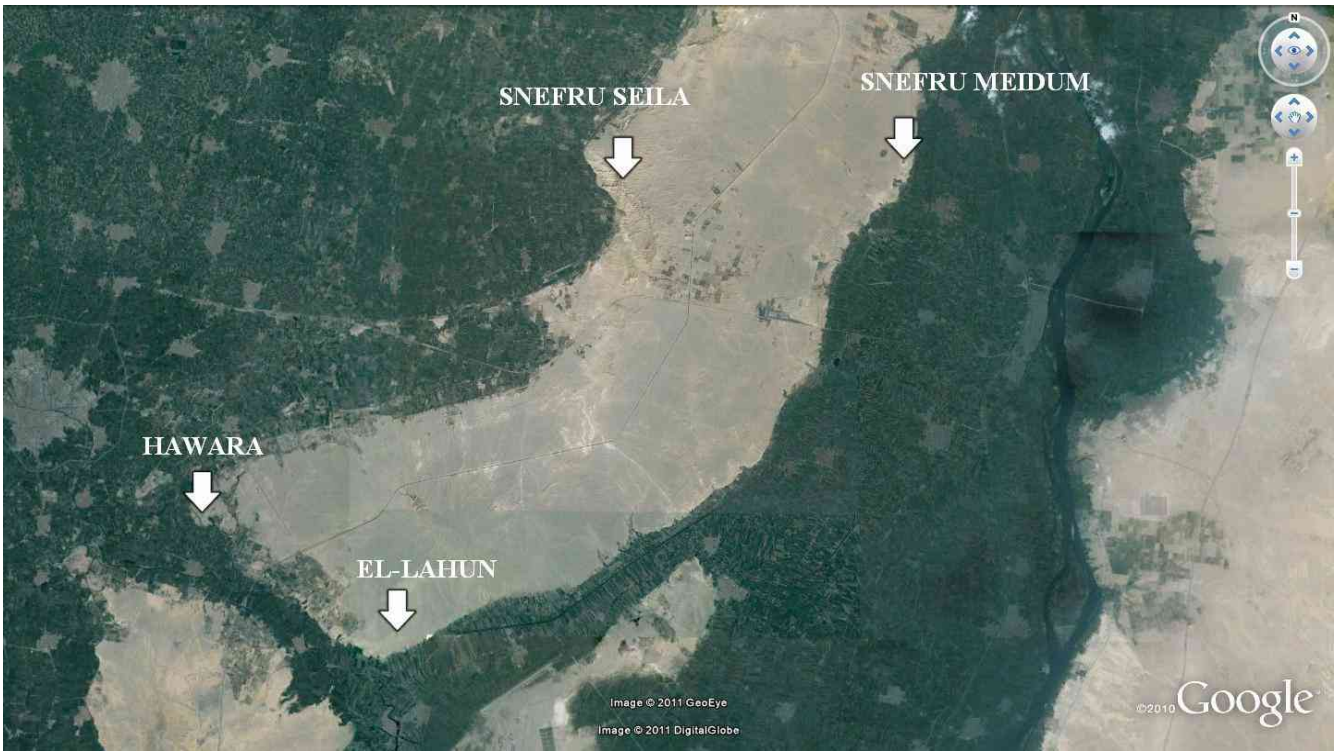

**Figure 5.** The position of the two royal pyramids in the Fayoum and that of Snefru pyramids at Meidum-Seila (image courtesy of Google Earth).

The ideal resemblance of the set of the Middle Kingdom projects in Dahshur to the Snefru project there can therefore be said to be complete with the completion of the Amenemhet III pyramid. As mentioned, perhaps also this pyramid was conceived only as a cenotaph. Amenemhet III had indeed another pyramid built for him, along the northern rim of the Fayoum oasis.

The reason for building royal pyramids in the Fayoum is probably due to the "interest" of Middle Kingdom pharaohs for this zone, which was subjected to intensive drainage works. However, if this may be enough to justify Senwostret II' choice of the mouth of the Fayoum (the modern village of Lahun) certainly it is *not* enough to understand why, given for grant the will of building in the area, Amenemhet III opted for Hawara. The two sites are 8.7 Kms far apart, and at Lahun there was plenty of space to build a new pyramid, in an already existing - "sanctified" - necropolis of a revered predecessor. Further, there were the "infrastructures" needed for pyramid construction: accessibility of materials and a huge pyramid's workers town. The explanation I have recently proposed for this riddle is that the choice of Hawara allowed the king to complete an ideal replica also of the second Snefru project, that of Meidum-Seila. First of all indeed, it must be observed that the two pyramids of El-Lahun and Hawara were inter-visible. Today it is difficult to understand the prominent role they played for any visitor of the Fayoum: the huge mass of El-Lahun is barely visible from Hawara, while the vice-versa is nearly impossible with the naked eye. However, in ancient times the two monuments clearly "spoke" with each other. As shown better by the satellite images, they actually stand as "paired sentinels" at the two corner ends of the strip of desert which is the prolongation to the south of the pyramid's fields ridge. Thus they played the role of "sentinels of power" in a pretty similar fashion to that played by Meidum and Seila. Still at the end of the 19th century Seila was inter-visible with Hawara; for example, Flinders Petrie [23] noticed their visual relationship and was led to think that the two monuments were built in the same period.

## 5. The sacred landscape at western Thebes

With the advent of the New Kingdom the capital becomes Thebes (modern Luxor). The first king of the New Kingdom, Amhose, built a funerary complex at Abydos, but the subsequent pharaohs choose to be buried in a rock-carved tomb on the west bank of the Nile at Thebes. These tombs were located in a Wadi which is the world-famous *Valley of the Kings*. The Valley is located behind the cliffs of the Deir el-Bahri bay. It was accessed in ancient times by a track over these cliffs, but also by a smooth route, probably used for the funeral of the king [24]. This route plays therefore the same role of the ceremonial pathway at Abydos and of the causeways of the pyramid's complexes; the role of "Valley Temples" is played by the monuments located at the edge of the cultivated land (see below).

The choice of the valley was with all probabilities influenced by symbolic criteria [7]. First of all, it has been repeatedly said in the Egyptological literature that its position behind the western horizon as seen from Thebes assimilated the king's death and rebirth with the solar cycle. Actually, such a statement can be made more quantitative by observing that the axis of the Karnak temple of Amon – by far the most important religious center in Egypt during the New Kingdom – passes quite precisely along the northern rim of the Deir el Bahri bay. The Karnak temple axis is oriented to the winter solstice sunrise to the south-east, the opposite orientation (which would be to the summer solstice sunset with a flat horizon) being "occupied" precisely be the hills which guard the entrance to the eastern branch of the Valley, where most of the tombs are located. The Valley by itself is signaled to the observer on the east bank by the terraces of the temple of Hatshepsut inside the bay.

Another symbolism embodied in the choice of the Valley is connected with the prominent peak called el-Qurn. The resemblance of this peak to a pyramid is obvious from any side, but becomes striking when the mountain is seen from the east, and therefore from the area of the temples, to a point that the present author doubts about the possibility that the natural pyramidal shape was willingly adapted by sculpting at least on this side. In any case, the peak in itself was not used to carve tombs, nor are orientation of tombs and temples aimed towards it, so that it has to be considered only as one important element of the landscape among the others. Again, it is a combination of elements – man made, and natural – which results in the construction of an extraordinary conceptual landscape.

The man-made elements are the so called *Temples of Millions of Years*. These temples were built at the border of the cultivated land, at the base of the cliffs surrounding El-Qurn. The southernmost is that of Ramesses III located at Medinet Habu, the northernmost one is that of Seti I, slightly to the north-east of Deir el Bahri. Practically each pharaoh buried in the Valley had his own temple constructed in the flatland, but the temples do not follow any recognizable chronological order on the ground. Generally speaking, it can be said that these monuments were devoted to the worship of the king, and in this sense they really are "funerary temples" and play a role similar to that of the early dynastic royal enclosures and the temples of the pyramids. However, their significance goes far beyond this, and to understand them correctly their relationship with the cult of Amun in the Karnak and Luxor temples, and again with the Pharaoh's cult in the same temples must be considered [25].

**Figure 6.** The funerary landscape of Ramesses III at Western Thebes (compare with Fig. 1). (1) Temple of Medinet Habu (2) Ancient route to the Valley (3) Royal tomb (4) Wadi (5) El-Qurn. The red line denotes the meridian (image courtesy of Google Earth)

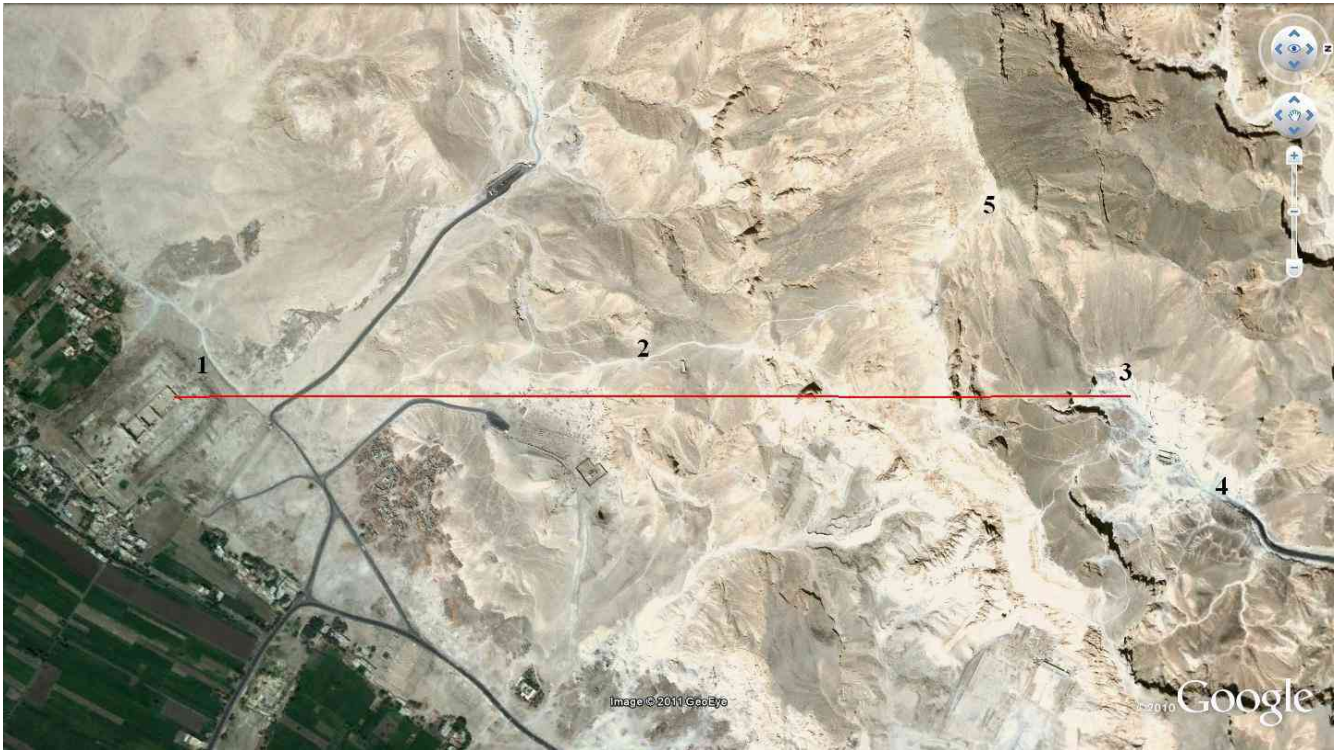

Since the valley is invisible from the flood plain, the El-Qurn peak is the representative element for the king's tomb in the funerary landscape at western Thebes. As far as the temples are concerned, once again, orientation issues play a relevant role. A case must be distinguished, namely that of two pharaohs who choose to orientate their temples to winter solstice sunrise. Not by chance these are the Pharaoh Queen Hatsheput, who had the necessity of claiming a sort of direct call from the Sun to legitimate her reign, and Amenhotep III, the father of the "heretic" Pharaoh Akhenaton. All the other temples belong to the inter-cardinal family, having orientations close to 135°. Why? No doubt, this orientation, being to the south of the winter solstice sunrise, has the practical consequence that the facade of the temple is fully illuminated by the climbing sun every day of the year. Also here it has been proposed that the orientation was obtained by determining celestial north trough the movement of circumpolar stars and then rotating this direction clockwise [14]; although this might be a functional orientation procedure, I do not believe its proposed origin – that of "mediating" between orientation to true north and orientation orthogonal to the Nile – to be feasible. In the ancient Egyptian mind "keeping order" was first of all "keeping tradition", sometimes in a nearly maniacal, time-traversing way. Actually the sacred space at western Thebes closely resembles the original sacred space at Abydos, a place which, meanwhile, had "evolved" as the true burying place of the most important God of Afterworld, Osiris. At Abydos, as we have seen, inter-cardinal orientation was the regional, strictly applied rule.

To show the resemblance between the two sacred landscapes I will use the most striking, neat example, warning the reader that, although the general directionality relationships are respected by all complexes, such a rigorous correspondence applies only to this case. The example is the funerary landscape created for Ramesses III. The landscape from north to south is so composed: the royal wadi, the tomb, the ancient (probably processional) route, the temple. Considering that the temple is due south of the tomb and oriented inter-cardinally (azimuth 137°), the complete analogy with the Khasekhemwy complex becomes evident.

**6. Amarna**

Amarna, in Middle Egypt, is the site once chosen by the pharaoh Akhenaten to found a new capital around 1346 BC [26]. In accordance with his religious revolution, based on the monotheistic cult of the solar disc Aten, the city was named Akhet-Aten that is "Horizon of Aten." At the death of the king Egypt rapidly came back to the old cults, and the newly founded town was abandoned. Amarna is, therefore, of extreme interest in order to study the conception of the sacred space that the "heretic" king elaborated in his new doctrine.

First of all, it should be noticed that the choice of a completely virgin site – directly chosen by the God Aten, accordingly to contemporary texts – stressed from the very beginning the neat rupture with the previous religion [4]. The sacred landscape at Amarna is an example of *consecrated landscape*, an environment which is ritually founded to assure its suitability for human beings to live. This process is typical, for instance, of the Etruscan-Roman world, were the foundation of a town was associated to a series of ritual acts (think e.g. to the myth of foundation of Rome). In the case of Amarna the ritual limitation of the sacred space took the peculiar form of the so called *boundary stelae*. The name is quite inappropriate because these "stelae" are monuments carved in the rocks of the cliffs surrounding the town on both the river's bank, carrying royal inscriptions and statues (sadly, most of them are destroyed today). These monuments "speak each other" trough visibility lines traceable between the two banks of the Nile. This sanctification of the whole urban landscape looks something of a novelty for the Egyptian world, where foundation rituals are very well known and documented in the archaeological records only in the case of temples and tombs. However, I believe that a sort of boundary of the sacred landscape at Thebes can perhaps be identified, at least to the north, by the mutual position of two temples located on the opposite banks of the Nile. These are the "Nest of Horus" on Thoth Hill, overlooking the valley of the Nile from the western bank, and the temple of Montu at Medamud, whose facade is oriented to Thoth Hill [21]. Both temples date back at least to the Middle Kingdom.

In spite of the novelty of many choices made, the pharaoh apparently decided to insert a series of understandable and well-established symbols and references for his own funerary landscape, namely the landscape associated with was meant to be in the future the cult of the deceased king (a cult which was never effective: the memory of the king was erased everywhere, he was probably never buried in Amarna and there is no certainty about the fate of his mummy). These symbols were, however, ordered and oriented in a "reverse" way. Indeed, the town extended on both banks, but the central cult area and the king's tomb were located into the *east* bank of the Nile. This is, of course, the first and fundamental rupture with the traditions: the deceased kings were usually associated with the "dying" sun to the west – and doomed to rebirth in the east. In a sense, Akhenaton is "already rebirth", his identification with the unique God, the King's father shining solar disc, occurs at the eastern horizon. It occurs, however, in the traditional way, i.e. referring to the "Akhet" concept which was effective at least since Khufu's reign.

**Figure 7.** The funerary landscape of king Akhenaten at Amarna.
(1) "Small" Temple of Aten (2) Wadi (3) Royal tomb.
The red line denotes the axis of the temple (image courtesy of Google Earth).

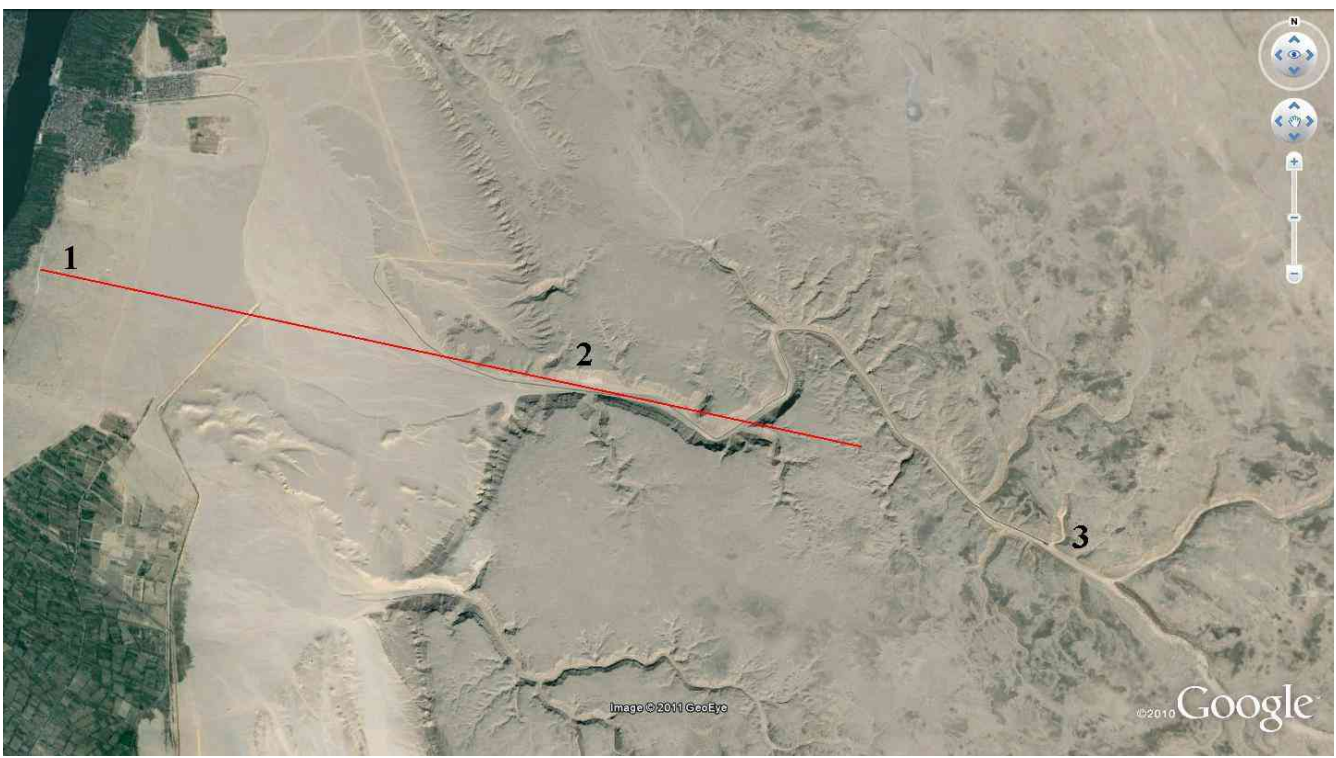

Indeed, in the cult area two temples are present, the so-called Great and Small Temple of the Aten. The presence of two parallel structures has never been explained satisfactorily. It is, however, likely that the "small" temple can be considered as the Amarna version of the pharaoh's funerary temple. The tomb is located in a Wadi, and the axis of the temple points to the mouth of it. Once again, we find the elements – funerary building, tomb, and wadi - already observed more than 1500 years before at Abydos. Here, the sun component is fundamental however, since the sun actually rises two times a year along the wadi and aligns to the axis of the temple.

**7. Discussion and conclusions.**

Conceptual, funerary landscapes in Egypt show a remarkable continuity in the use of symbols and in the integration between natural and man-built features. A key element to understand them is the analysis of *directionality*. It appears prominently into two ways: the arrangement in which elements of the landscape follow each other, and the orientation of single buildings and tombs. Taken together, these features play a key role in giving "order" to the landscape. Actually, they govern the architectural choices in accordance with the idea of "cosmic" order. In the present paper, I have tried to show how comparing satellite image with local surveys and using simple instruments for tracing visibility lines may help in understanding these connections. Indeed, although messages of power - alluding, for instance, to divine rights of kingship - were meant to be clear and obvious in ancient times, they may be lost, or forgotten, today.

It is, I believe, an exciting experience to see how modern technologies can help us in unraveling such messages from the mists (both real, and metaphoric) of such an ancient past.

**References**

1. Ruggles C. L. N. (2005). *Ancient Astronomy: An Encyclopedia of Cosmologies and Myth*. London: ABC-CLIO.
2. Magli, G. (2009) *Mysteries and Discoveries of Archaeoastronomy*, Springer-Verlag, NY
3. Flannery, K., Marcus J. (1996) *Cognitive Archaeology* In *Contemporary Archaeology in Theory: A Reader (Social Archaeology)* R. W. Preucel and I. Hodder eds. Wiley-Blackwell NY
4. Richards, J.E., (1999) Conceptual Landscapes in the Egyptian Nile Valley. In W.Ashmore and B. Knapp, Eds. *Archaeologies of Landscape: Contemporary Perspectives*.London: Blackwell 83-100.
5. Lehner, M., (1985) A contextual approach to the Giza pyramids, Archiv fur Orientf.. 31, 136-158
6. Jeffreys, D. (1998) The topography of Heliopolis and Memphis: some cognitive aspects, Beitrage zur Kulturgeschichte Ägyptens, Rainer Stadelmann gewidmet (Mainz) 63-71.
7. Romer, John (1981), *Valley of the Kings*; New York, NY: Henry Holt and Company.
8. Magli, G. (2010) Topography, astronomy and dynastic history in the alignments of the pyramid fields of the Old Kingdom; Mediterranean Archaeology and Archaeometry 10, 59-74.
9. Magli, G. (2010) The Cosmic Landscape in the Age of the Pyramids. Journal of Cosmology 9, 3132-3144
10. Magli, G. (2010) Archaeoastronomy and Archaeo-Topography as Tools in the Search for a Missing Egyptian Pyramid. PalArch's Journal of Archaeology of Egypt/Egyptology, 7(5)
11. Potere, D. (2008) Horizontal Positional Accuracy of Google Earth's High-Resolution Imagery. Archive Sensors 8: 7973-7981.
12. Shaw, Ian (2000). The Oxford history of ancient Egypt. Oxford University Press
13. O'Connor, D. (2009) Abydos: Egypt's first pharaohs and the cult of Osiris. Thames & Hudson, London.
14. Belmonte, J.A., Shaltout, M. , and Fekri, M. (2009) On the Orientation of Ancient Egyptian Temples: (4) Epilogue in Serabit el Khadim and Overview J.H.A. 39
15. Lehner, M. (1999) The complete pyramids, Thames and Hudson, London.
16. Verner, M. (2002) The Pyramids: The Mystery, Culture, and Science of Egypt's Great Monuments Grove Press

17. Hawass, Z. (1993) The Great Sphinx at Giza: Date and Function In Sesto Congresso Internazionale di Egittologia. Ed. G.M. Zaccone and T. Ricardi di Netro, pp. 177–195.Turin.

18. Magli, G. (2009) Akhet Khufu: Archaeo-astronomical Hints at a Common Project of the Two Main Pyramids of Giza, Egypt. NNJ- Architecture and Mathematics 11, 35-50.

19. Silverman, D, Simpson, W.K., and Wegner, J. (eds.) *Archaism and Innovation: Studies in the Culture of Middle Kingdom Egypt* Yale University and University of Pennsylvania Museum of Archaeology and Anthropology press.

20. Magli, G. (2010) A cognitive approach to the topography of the 12th dynasty pyramids. Preprint arxiv.org/abs/1011.2122

21. Shaltout, M., Belmonte, J.A. and Fekri, M. (2007) On the Orientation of Ancient Egyptian Temples: (3) Key Points in Lower Egypt and Siwa Oasis. Part II. Journal for the History of Astronomy 38: 413-422.

22. Belmonte, J. (2009) *The Egyptian calendar: keeping Maat on earth* In *In Search Of Cosmic Order - selected Essays on Egyptian Archaeoastronomy* Edited by J.A. Belmonte and M. Shaltout, Supreme Council of Antiquities Press, Cairo.

23. Petrie, F. (1890 ) *Kahun, Gurob, and Hawara.* Eisenlohr collection in Egyptology and assyriology, Cornell University press.

24. Reeves, C.N., Wilkinson, R.H. (1996) *The Complete Valley of the Kings.* Thames & Hudson, London.

25. Arnold D., Haeny G., Bell L., Bjerre Finnestad, B., Shafer B.F. (1997) *Temples of Ancient Egypt.* Cornell University Press

26. Redford, D. (1987) *Akhenaten, the heretic king*, Princeton University Press, Princeton.